\documentclass[pre,twocolumn,superscriptaddress,showpacs,preprintnumbers,amsmath,amssymb]{revtex4-1}

\usepackage{graphicx}
\usepackage{dcolumn}
\usepackage{bm}
\usepackage{multirow}

\begin{document}

\bibliographystyle{apsrev4-1} 


\title{Equivalence condition for the canonical and microcanonical ensembles \\in coupled spin systems}

\author{Wenxian Zhang}
\affiliation{Key Laboratory of Micro and Nano Photonic Structures, Department of Optical Science and Engineering, Fudan University, Shanghai 200433, China}
\affiliation{Advanced Science Institute, RIKEN, Wako-shi, Saitama 351-0198, Japan}

\author{C. P. Sun}
\affiliation{Advanced Science Institute, RIKEN, Wako-shi, Saitama 351-0198, Japan}
\affiliation{Institute of Theoretical Physics, Chinese Academy of Science, Beijing 100080, China}

\author{Franco Nori}
\affiliation{Advanced Science Institute, RIKEN, Wako-shi, Saitama 351-0198, Japan}
\affiliation{Department of Physics, The University of Michigan, Ann Arbor, Michigan 48109-1040, USA}

\date{\today}

\begin{abstract}
It is typically assumed, without justification, that a weak coupling between a system and a bath is a necessary condition for the equivalence of a canonical ensemble and a microcanonical ensemble. For instance, in a canonical ensemble, temperature emerges if the system and the bath are uncoupled or weakly coupled. We investigate the validity region of this weak coupling approximation, using a coupled composite-spin system. Our results show that the spin coupling strength can be as large as the level spacing of the system, indicating that the weak coupling approximation has a much wider region of validity than usually expected.
\end{abstract}


\maketitle

\section{Introduction}

It is of fundamental importance to revisit statistical mechanics, based on principles of quantum mechanics (see e.g.,~\cite{Landau58, Tasaki98, Goldstein06, Popescu06, Lloyd06, Quan06, Quan07, Maruyama09}). Starting from a microcanonical ensemble, the concept of temperature emerges from a canonical ensemble in the limit of weak coupling between a system and a bath. This is valid both for classical distributions on phase space and for quantum density matrices (see, e.g.,~\cite{Landau58, Tasaki98, Goldstein06}). A general composite system made of a central system and a bath can be described by
\begin{eqnarray}
\label{eq:h}
H &=& H_S + H_B + H_{SB},
\end{eqnarray}
with $H_S$ the system Hamiltonian, $H_B$ the bath Hamiltonian, and $H_{SB}$ the system-bath coupling. In order to obtain the temperature, the small term $H_{SB}$ is usually neglected.

In quantum mechanics, the density matrix of a thermal equilibrium state in a microcanonical ensemble with energy between $[E,E+\delta]$ ($\delta \ll E$ but much larger than the level spacing of the bath) is described by
\begin{eqnarray}
\rho^C &=& \frac{1}{d} {\cal P}_C
\end{eqnarray}
where $d$ is the dimension of the subspace with total energy in the interval  $[E,E+\delta]$, and ${\cal P}_C$ is the projection onto that subspace. In the weak coupling limit, $H_{SB} \rightarrow 0$, the canonical ensemble distribution is derived from the microcanonical ensemble by following the standard procedure (see, e.g.,~\cite{Landau58, Goldstein06})
\begin{eqnarray}
\label{eq:ce}
\rho^S &\equiv & {\rm Tr}_B (\rho^C) \nonumber \\
    &=& \frac{e^{-\beta H_S}}{Z}
\end{eqnarray}
with the partition function $Z={\rm Tr}[\exp(-\beta H_S)]$ and the inverse temperature $\beta = 1/T$. A much stronger statement, called canonical typicality, has also been made, which states that an arbitrary pure random state in the microcanonical subspace is enough to derive the above canonical ensemble result (see, e.g.,~\cite{Popescu06, Goldstein06}). Several works have followed this direction (see, e.g.,~\cite*{Rigol07, Yuan07, Cramer08a, Cramer08b, Camalet08, Reimann08, Rigol08, Gartia09, Vine09, Yuan09, Linden09, Cho10, Schroder10}). However, the meaning of ``weak coupling" has been invoked without any justification, except in the papers by Dong {\it et al.}~\cite{Dong07} and Reimann~\cite{Reimann10}.

In this paper, we investigate the effect of coupling between the system and the bath and justify the validity of the weak-coupling approximation in coupled-spin systems. These spin systems are very different from the harmonic oscillators in Ref.~\cite{Dong07}. This paper is organized as follows. We describe the model of a coupled-spin composite system in Sec.~\ref{sec:model} and present our numerical results in Sec.~\ref{sec:results}. A brief discussion and a conclusion are given in Sec.~\ref{sec:discon}.

\section{\label{sec:model} Coupled spin systems}

\begin{figure}
\includegraphics[width=3in]{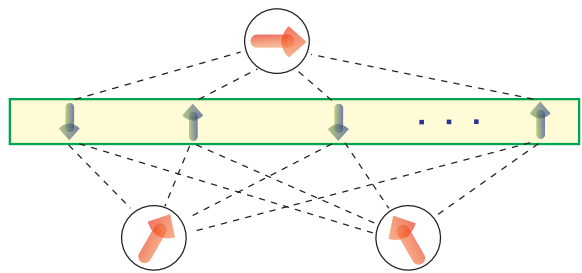}
\caption{\label{fig:model}(Color online) Schematic diagram of a spin system (red arrows with circles) coupled to a spin bath (blue arrows). The dashed lines denote the coupling between them.}
\end{figure}

We consider a model for many spin-1/2 particles (see Fig.~\ref{fig:model}). Some of these spins are labeled as ``system spins", denoted by ${\mathbf S}_m$, with $m = 1, 2, \dots,M$. Others are labeled as ``bath spins", denoted by ${\mathbf I}_k$, with $k = 1, 2, \dots, K$. This composite system is described by the Hamiltonian in Eq.~(\ref{eq:h}), where
\begin{eqnarray}
\label{eq:hc}
H_S & = & \sum_{m=1}^M \Omega_m\, \hat S_{mz}, \nonumber \\
H_B & = & \sum_{k=1}^K \omega_k\, \hat I_{kz}, \\
H_{SB} & = & \sum_{m=1}^M \sum_{k=1}^K A_{km}\, \hat S_{mz}\, \hat I_{kx} \nonumber
\end{eqnarray}
with $A_{km}$ describing the coupling between the system spin $\mathbf{S}_m$ and the bath spin $\mathbf{I}_k$. Here, $\Omega_m$ and $\omega_k$ are the Zeeman splittings of the $m$th system spin and the $k$th bath spin, respectively. We have set $\hbar = 1$ and $k_B=1$ for convenience. Note that the system and the bath have dimensions of $2^M$ and $2^K$, respectively. The system is only experiencing dephasing, because of the commutation between the system $H_S$ and the coupling $H_{SB}$, $\left[H_S, H_{SB}\right] = 0$.

The above model [Eqs.~(\ref{eq:h}) and (\ref{eq:hc})] is exactly solvable. Without the system-bath coupling term, the total energy $E^T$ is the sum of the system energy $E^S$ and the bath energy $E^B$, i.e., $E^T = E^S + E^B$. For an eigenstate, the eigenvalue of $H$ is
\begin{eqnarray}
E_{ij} &=& E_i^S + E_{j}^B \nonumber \\
    &=& \sum_{m=1}^M \Omega_m\, s_{mz} + \sum_{k=1}^K \omega_k\, I_{kz},
\end{eqnarray}
where $s_{mz}$ and $I_{kz}$ are the eigenvalues of $\hat S_{mz}$ and $\hat I_{kz}$, respectively. The eigenbasis are the usual computational basis. Here, $i$ and $j$ label the eigenstates, and range from 1 to $2^M$ and from 1 to $2^K$, respectively.

With the system-bath coupling term, the eigenvalue of an eigenstate becomes
\begin{eqnarray}
E_{i,j_\pm} & = & \sum_{m=1}^M \Omega_m\, s_{mz} \pm \left[ {1\over 2} \sum_{k=1}^K \sqrt{B_k^2 + \omega_k^2} \right]
\end{eqnarray}
with $B_k = \sum_{m=1}^M A_{km}\, s_{mz}$ and the corresponding eigenvector is
\begin{equation}
|\psi_{i,j_\pm}\rangle = \bigotimes_{m=1}^M |s_{mz} \rangle \bigotimes_{k=1}^K |I_{k\pm}\rangle,
\end{equation}
where
\begin{eqnarray}
\left|I_{k+}\right\rangle &=& \cos (\theta_k/2) \left|I_{kz}={1\over 2}\right\rangle + \sin(\theta_k/2) \left|I_{kz}=-{1\over 2} \right\rangle, \nonumber \\
\left|I_{k-}\right\rangle &=& \sin (\theta_k/2) \left|I_{kz}={1\over 2}\right\rangle - \cos(\theta_k/2) \left|I_{kz}=-{1\over 2} \right\rangle \nonumber
\end{eqnarray}
with $\cos\theta_k = B_k / \sqrt{B_k^2 + \omega_k^2}$ and $\sin\theta_k = \omega_k / \sqrt{B_k^2 + \omega_k^2}$\;. Similarly, $i$ (from 1 to $2^M$) and $j_{\pm}$ (from 1 to $2^{K-1}$) label the eigenstates.

\section{\label{sec:results} Numerical results}

In our calculations, we choose a set of random numbers $\Omega_m \in [0, 0.4]$, $\omega_k \in [0,1]$, $A_{km} \in [0,0.037]$. The number of system spins and bath spins are $M=3$ and $K=18$, respectively. For example, here we set the energy shell $[E,E+\delta]$ by choosing $E=-1.722$ and $\delta = 0.0350$. We count the total number of composite-system eigenstates whose energy is within the given energy shell $[E,E+\delta]$ and partition this number according to different system eigenstates into $2^M$ numbers. In this way, we obtain the probabilities $P_i$ ($i=1,2,\cdots,8$ for $M=3$). We adjust the coupling strength by multiplying an integer $n$ (e.g., $n=0,3,10,30$) for all $A_{km}$. Thus, $n=0$ corresponds to the case without coupling, $n=3$ weak coupling, and larger $n$ denotes stronger coupling.

\begin{figure}
\includegraphics[width=3.5in]{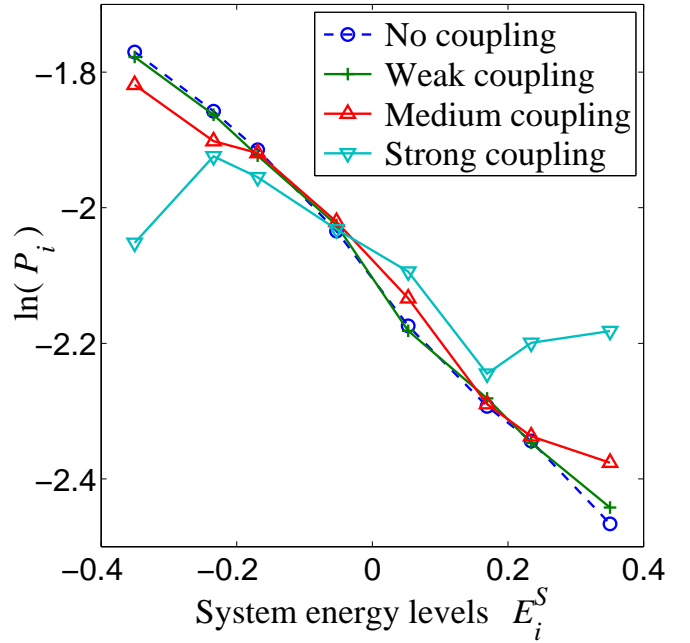}
\caption{\label{fig:pe}(Color online) Canonical relation between $\ln(P_i)$ and the system energy $E_i^S$ for various system-bath coupling strengths: $n=0$ (no coupling, blue dashed line with circles), $n=3$ (weak coupling, green solid line with crosses), $n=10$ (medium coupling, red solid line with upper triangles), and $n=30$ (strong coupling, cyan solid line with down triangles). Here $P_i$ is the probability that the system is in the $i$th eigenstate with energy $E_i^S$. For weak and medium couplings ($n=0,3,10$), the curves are almost linear. See also Table~\ref{tbl:br}.}
\end{figure}

\begin{table}
\caption{\label{tbl:br} 
Inverse temperatures $\beta$ extracted by fitting the slope of the curves in Fig.~\ref{fig:pe}, and the corresponding linear correlation coefficient $r$ of each fit.}
\begin{tabular}{c|cccc}
\hline
\hline
Coupling strength  $n$ & 0 & 3 & 10 & 30 \\
\hline
Inverse temperature  $\beta$ & 1.027 & 0.993 & 0.875 & 0.388 \\
Linear correlation coefficient  $r$ & -0.998 & -0.998 & -0.990 & -0.815 \\
\hline
\hline
\end{tabular}
\end{table}

We plot in Fig.~\ref{fig:pe} the relation between $\ln(P_i)$ and $E_i^S$. One would expect a {\it linear} relationship {\it if} the canonical ensemble [Eq.~(\ref{eq:ce})] is equivalent to the microcanonical ensemble. We clearly see a linear relationship in the cases of no coupling ($n=0$) and weak coupling ($n=3$) as well as deviations from a linear relationship in the cases of medium coupling ($n=10$) and specially for strong coupling ($n=30$). The stronger the coupling between the system and the bath spins, the larger the deviation from a straight line. By fitting these data with straight lines, we can obtain the inverse temperature $\beta$ from each data set, i.e., the negative slope of the fitted line. We also obtain the linear correlation coefficient $r$, which shows how good the linear relationship is, where $|r|=1$ corresponds to a straight line. The values of $\beta$ and $r$ are listed in Table~\ref{tbl:br} for the four cases we present in Fig.~\ref{fig:pe}. Again, we find from these values better linear relationships in the cases of no coupling and weak coupling.

From the above numerical results, we observe that the canonical ensemble distribution is valid in the cases of no coupling and weak coupling,  and approximately valid for medium coupling. By comparing the level-spacing $s\approx 0.1$ of the system and the typical coupling strength of the medium-coupling-case, $A_{km}\lesssim 0.37$, we find that the canonical ensemble distribution is a good approximation if the coupling strength is less or of the order of the level spacing, $A_{km} \lesssim s$. In other words, {\it the canonical ensemble is equivalent to the microcanonical ensemble if the typical coupling terms are less or equal to the level spacing of the system.} However, the equivalence between the two ensembles is broken once $A_{km} \gg s$, as shown in the strong coupling case $n=30$ in Fig.~\ref{fig:pe}(d). Compared to Tasaki's estimation $A_{km}\ll s$~\cite{Tasaki98}, our results give a much wider range, which significantly extends the previous smallness requirement on the system-bath coupling.

\begin{figure}
\includegraphics[width=3.25in]{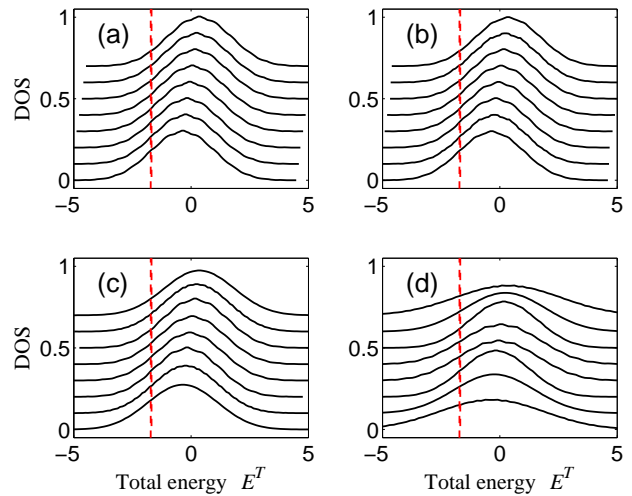}
\caption{\label{fig:dos}(Color online) Normalized density of states for system-bath coupling strength $n=0$ (a), 3 (b), 10 (c), 30 (d). The red lines show the energy range $[E,E+\delta]$ for small $\delta \ll E$. Here, $E=-1.722$.}
\end{figure}

Next we consider how $\beta$ depends on the composite system's parameters. We plot in Fig.~\ref{fig:dos} the normalized density of states (DOS) distribution for each system eigenstate. For clarity, we shift each curve up $0.1$ in all panels. As expected for spin systems, one clearly sees Gaussian distributions for the density of states. It is especially interesting to note that the widths of the Gaussian distributions are almost identical in panels~\ref{fig:dos}(a)-(c), where one finds that the canonical-ensemble distribution is valid. We list in Table~\ref{tbl:w} the width of the Gaussian fit of the curves in Fig.~\ref{fig:dos}.

\begin{table}
\caption{\label{tbl:w} 
The width of the Gaussian fit of the curves, $\sigma_0$, in Fig.~\ref{fig:dos}. Due to the symmetry, only the 4 lowest curves are listed.}
\begin{tabular}{c|cccc}
\hline
\hline
\multirow{2}{*}{System-bath coupling strength} & \multicolumn{4}{c}{Level index} \\
& 1 & 2 & 3 & 4 \\
\hline
$n=0$ & 1.89 & 1.89 & 1.89 & 1.89 \\
$n=3$ & 1.90 & 1.89 & 1.89 & 1.89 \\
$n=10$ & 2.06 & 1.95 & 1.90 & 1.94 \\
$n=30$ & 3.11 & 2.37 & 2.00 & 2.34 \\
\hline
\hline
\end{tabular}
\end{table}

\subsection{Weak-coupling limit}

From Fig.~\ref{fig:dos}(a)-(c), it is clear that the bath-level distribution for each system eigenstate is the same, except for a shifted peak position. In the ``weak" coupling limit, let us assume that the probability of the system to be in the $i$th eigenstate (with the total energy $E$) is
\begin{eqnarray}
P_i(E) &=& \frac{1}{\sigma_0 \sqrt{\pi}} \exp\left[-\frac{(E-E_i)^2}{\sigma_0^2}\right]\; ,
\end{eqnarray}
 where $\sigma_0$ is the width of the bath spectrum. Within the energy shell $[E,E+\delta]$, the probability becomes
\begin{eqnarray}
P_i &=& \int_{E}^{E+\delta} P_i(x) dx \simeq P_i(E) \; \delta.
\end{eqnarray}
For two system eigenstates, the ratio of probabilities is
\begin{eqnarray}
\label{eq:p}
\frac{P_i}{P_{i'}} &=& \exp\left[-\,\frac{(E-E_i)^2-(E-E_{i'})^2}{\sigma_0^2}\right] \nonumber \\
&\approx & \frac{\exp(-\beta E_i)}{\exp(-\beta E_{i'})}
\end{eqnarray}
where $\beta \approx -2E/\sigma_0^2$, if $E \gg E_i, E_{i'}$. Using Eq.~(\ref{eq:p}), we can estimate the inverse temperature $\beta$ for a given $E$ and a bath spectrum width $\sigma_0$. For the zero-coupling case, where the bath-spectrum widths are the same for all states, we find $\beta = 0.964$ in the cases shown in Fig.~\ref{fig:pe}(a)-(c). This value of $\beta$ is in good agreement with the first three values shown in Table~\ref{tbl:br}.

\subsection{Strong-coupling limit}

In the strong-coupling limit, the assumption of $\sigma_0$ being the same for all system eigenstates does not hold, as shown in Fig.~\ref{fig:dos}(d). In this limit, it is impossible to obtain the (inverse) temperature, since the canonical ensemble and the microcanonical ensemble are not equivalent. From the features of the density of states, we thus conclude that the equivalence between a canonical ensemble and a microcanonical ensemble happens only when the system-bath coupling is ``weak", so that the coupling does not significantly change the spectrum properties, e.g., the spectrum width $\sigma_0$.

We would like to remark that we may obtain a negative inverse temperature ($\beta < 0$) if the total energy of the composite system is positive ($E>0$), according to the evaluation of the density of states.

\begin{figure}
\includegraphics[width=3.5in]{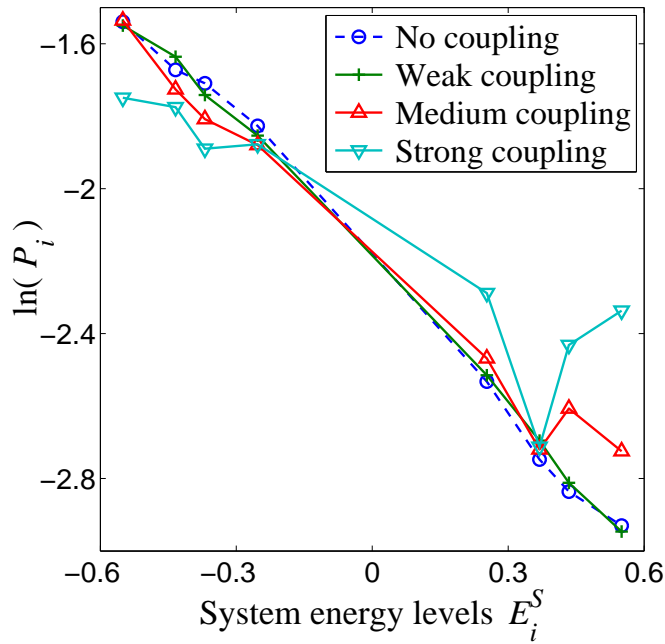}
\caption{\label{fig:pec}(Color online) Same as Fig.~\ref{fig:pe} except that the system level distribution is now uneven. Note that this implies that an uneven level spacing distribution is equally sensitive to the system-bath coupling strength.}
\end{figure}

\begin{table}
\caption{\label{tbl:brc} 
Same as Table~\ref{tbl:br} except that these values are extracted from Fig.~\ref{fig:pec}.}
\begin{tabular}{c|cccc}
\hline
\hline
Coupling strength  $n$ & 0 & 3 & 10 & 30 \\
\hline
Inverse temperature  $\beta$ & 1.325 & 1.300 & 1.101 & 0.734 \\
Linear correlation coefficient  $r$ & -0.998 & -0.999 & -0.992 & -0.915 \\
\hline
\hline
\end{tabular}
\end{table}

We now tentatively conjecture that a system with evenly-distributed levels is easier to thermalize, and thus more sensitive to the coupling strength. With the help of our numerics, we check below this conjecture by comparing the previous almost-even level distribution case with the following uneven one. We find that this conjecture is invalid.

\begin{figure}
\includegraphics[width=3.25in]{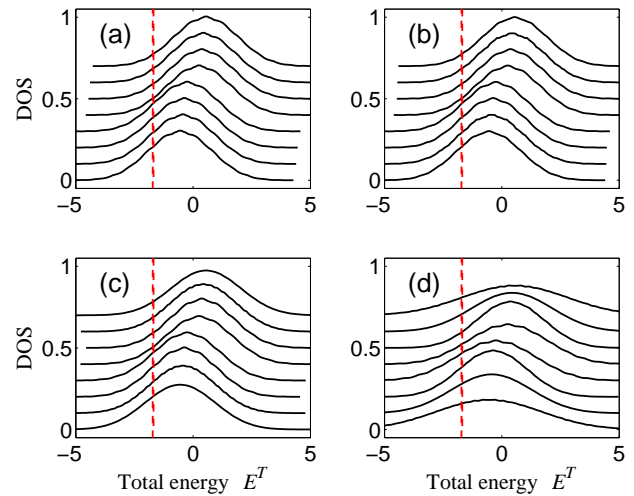}
\caption{\label{fig:dosc}(Color online) Same as Fig.~\ref{fig:dos} except for a system with uneven level distribution. The total energy (the red vertical dashed line) is $E=-2.214$. Note that the DOS is not significantly affected by the uneven distribution of the system energy levels.}
\end{figure}

We keep all the parameters the same as before except $\Omega_m$. We then choose another set of $\Omega_m$, which generates unevenly-distributed energy levels for the system (see Fig.~\ref{fig:pec}). The results are presented in Figs.~\ref{fig:pec} and \ref{fig:dosc}, and in Tables~\ref{tbl:brc} and \ref{tbl:wc}. Comparing the two cases, especially the linear correlation coefficients $r$ in Table~\ref{tbl:br} and \ref{tbl:brc}, we do not find strong evidence to support the above conjecture that systems with evenly-distributed levels are more sensitive to the coupling strength between the system and the bath.

\begin{table}
\caption{\label{tbl:wc} 
The width of the Gaussian fit of the 4 lowest curves in Fig.~\ref{fig:dosc}.}
\begin{tabular}{c|cccc}
\hline
\hline
\multirow{2}{*}{System-bath coupling strength} & \multicolumn{4}{c}{Level index} \\
& 1 & 2 & 3 & 4 \\
\hline
$n=0$ & 1.89 & 1.89 & 1.89 & 1.89 \\
$n=3$ & 1.90 & 1.89 & 1.89 & 1.89 \\
$n=10$ & 2.06 & 1.95 & 1.90 & 1.94 \\
$n=30$ & 3.11 & 2.38 & 2.00 & 2.34 \\
\hline
\hline
\end{tabular}
\end{table}

\section{\label{sec:discon} Discussion and conclusion}

The composite system we calculate is finite, with $M=3$ spins for the system and $K=18$ spins for the bath. The total number of eigenstates is $2^{21} \approx 2\times 10^6$. But the number of eigenstates in the energy shell $[E, E+\delta]$ is not very large ($\sim 10^4$) for the parameters we choose. From the numerical point of view, the relative statistical error of the data we present in Figs.~\ref{fig:pe} and \ref{fig:pec} is of the order of $1\%$. Choosing a larger composite system would reduce the statistical error, with the price of more computation time. But we believe that our main conclusions would still hold qualitatively.

In summary, we investigate the effect (on the canonical ensemble distribution) of the the coupling between a spin system and a spin bath. We find that the canonical ensemble distribution still holds even if the typical coupling strength has the same order of magnitude as the system's level spacing. This is much larger than what was expected~\cite{Tasaki98}. In addition, we observe that the inverse temperature $\beta$ for a composite spin system can be expressed as $\beta = -2E/\sigma_0\,$, with $E$ being the energy in the microcanonical ensemble and $\sigma_0$ the width of the density of states in the system's eigenstates.

\section{acknowledgments}

WZ acknowledges support by the 973 Program (Grant No. 2009CB929300) and the National Natural Science Foundation of China (Grant No. 10904017). FN acknowledges partial support from the Laboratory of Physical Sciences, National Security Agency, Army Research Office, National Science Foundation grant No. 0726909, JSPS-RFBR contract No. 09-02-92114, Grant-in-Aid for Scientific Research (S), MEXT Kakenhi on Quantum Cybernetics, and Funding Program for Innovative R\&D on S\&T (FIRST).


\begin{thebibliography}{23}%
\makeatletter
\providecommand \@ifxundefined [1]{%
 \@ifx{#1\undefined}
}%
\providecommand \@ifnum [1]{%
 \ifnum #1\expandafter \@firstoftwo
 \else \expandafter \@secondoftwo
 \fi
}%
\providecommand \@ifx [1]{%
 \ifx #1\expandafter \@firstoftwo
 \else \expandafter \@secondoftwo
 \fi
}%
\providecommand \natexlab [1]{#1}%
\providecommand \enquote  [1]{``#1''}%
\providecommand \bibnamefont  [1]{#1}%
\providecommand \bibfnamefont [1]{#1}%
\providecommand \citenamefont [1]{#1}%
\providecommand \href@noop [0]{\@secondoftwo}%
\providecommand \href [0]{\begingroup \@sanitize@url \@href}%
\providecommand \@href[1]{\@@startlink{#1}\@@href}%
\providecommand \@@href[1]{\endgroup#1\@@endlink}%
\providecommand \@sanitize@url [0]{\catcode `\\12\catcode `\$12\catcode
  `\&12\catcode `\#12\catcode `\^12\catcode `\_12\catcode `\%12\relax}%
\providecommand \@@startlink[1]{}%
\providecommand \@@endlink[0]{}%
\providecommand \url  [0]{\begingroup\@sanitize@url \@url }%
\providecommand \@url [1]{\endgroup\@href {#1}{\urlprefix }}%
\providecommand \urlprefix  [0]{URL }%
\providecommand \Eprint [0]{\href }%
\@ifxundefined \urlstyle {%
  \providecommand \doi  [0]{\begingroup \@sanitize@url \@doi}%
  \providecommand \@doi [1]{\endgroup \@@startlink {\doibase
  #1}doi:\discretionary {}{}{}#1\@@endlink }%
}{%
  \providecommand \doi  [0]{doi:\discretionary{}{}{}\begingroup
  \urlstyle{rm}\Url }%
}%
\providecommand \doibase [0]{http://dx.doi.org/}%
\providecommand \Doi [0]{\begingroup \@sanitize@url \@Doi }%
\providecommand \@Doi  [1]{\endgroup\@@startlink{\doibase#1}\@@Doi}%
\providecommand \@@Doi [1]{#1\@@endlink}%
\providecommand \selectlanguage [0]{\@gobble}%
\providecommand \bibinfo  [0]{\@secondoftwo}%
\providecommand \bibfield  [0]{\@secondoftwo}%
\providecommand \translation [1]{[#1]}%
\providecommand \BibitemOpen [0]{}%
\providecommand \bibitemStop [0]{}%
\providecommand \bibitemNoStop [0]{.\EOS\space}%
\providecommand \EOS [0]{\spacefactor3000\relax}%
\providecommand \BibitemShut  [1]{\csname bibitem#1\endcsname}%
\bibitem [{\citenamefont {Landau}\ and\ \citenamefont
  {Lifshitz}(1958)}]{Landau58}%
  \BibitemOpen
  \bibfield  {author} {\bibinfo {author} {\bibfnamefont {L.~D.}\ \bibnamefont
  {Landau}}\ and\ \bibinfo {author} {\bibfnamefont {E.~M.}\ \bibnamefont
  {Lifshitz}},\ }\href@noop {} {\emph {\bibinfo {title} {Statistical
  Physics}}}\ (\bibinfo  {publisher} {Pergamon, London},\ \bibinfo {year}
  {1958})\BibitemShut {NoStop}%
\bibitem [{\citenamefont {Tasaki}(1998)}]{Tasaki98}%
  \BibitemOpen
  \bibfield  {author} {\bibinfo {author} {\bibfnamefont {H.}~\bibnamefont
  {Tasaki}},\ }\Doi {10.1103/PhysRevLett.80.1373} {\bibfield  {journal}
  {\bibinfo  {journal} {Phys. Rev. Lett.},\ }\textbf {\bibinfo {volume} {80}},\
  \bibinfo {pages} {1373} (\bibinfo {year} {1998})}\BibitemShut {NoStop}%
\bibitem [{\citenamefont {Goldstein}\ \emph {et~al.}(2006)\citenamefont
  {Goldstein}, \citenamefont {Lebowitz}, \citenamefont {Tumulka},\ and\
  \citenamefont {Zangh\`\i{}}}]{Goldstein06}%
  \BibitemOpen
  \bibfield  {author} {\bibinfo {author} {\bibfnamefont {S.}~\bibnamefont
  {Goldstein}}, \bibinfo {author} {\bibfnamefont {J.~L.}\ \bibnamefont
  {Lebowitz}}, \bibinfo {author} {\bibfnamefont {R.}~\bibnamefont {Tumulka}}, \
  and\ \bibinfo {author} {\bibfnamefont {N.}~\bibnamefont {Zangh\`\i{}}},\
  }\Doi {10.1103/PhysRevLett.96.050403} {\bibfield  {journal} {\bibinfo
  {journal} {Phys. Rev. Lett.},\ }\textbf {\bibinfo {volume} {96}},\ \bibinfo
  {pages} {050403} (\bibinfo {year} {2006})}\BibitemShut {NoStop}%
\bibitem [{\citenamefont {Popescu}\ \emph {et~al.}(2006)\citenamefont
  {Popescu}, \citenamefont {Short},\ and\ \citenamefont {Winter}}]{Popescu06}%
  \BibitemOpen
  \bibfield  {author} {\bibinfo {author} {\bibfnamefont {S.}~\bibnamefont
  {Popescu}}, \bibinfo {author} {\bibfnamefont {A.~J.}\ \bibnamefont {Short}},
  \ and\ \bibinfo {author} {\bibfnamefont {A.}~\bibnamefont {Winter}},\
  }\href@noop {} {\bibfield  {journal} {\bibinfo  {journal} {Nat. Phys.},\
  }\textbf {\bibinfo {volume} {2}},\ \bibinfo {pages} {754} (\bibinfo {year}
  {2006})}\BibitemShut {NoStop}%
\bibitem [{\citenamefont {Lloyd}(2006)}]{Lloyd06}%
  \BibitemOpen
  \bibfield  {author} {\bibinfo {author} {\bibfnamefont {S.}~\bibnamefont
  {Lloyd}},\ }\Doi {10.1038/nphys456} {\bibfield  {journal} {\bibinfo
  {journal} {Nat. Phys.},\ }\textbf {\bibinfo {volume} {2}},\ \bibinfo {pages}
  {727} (\bibinfo {year} {2006})}\BibitemShut {NoStop}%
\bibitem [{\citenamefont {Quan}\ \emph {et~al.}(2006)\citenamefont {Quan},
  \citenamefont {Wang}, \citenamefont {Liu}, \citenamefont {Sun},\ and\
  \citenamefont {Nori}}]{Quan06}%
  \BibitemOpen
  \bibfield  {author} {\bibinfo {author} {\bibfnamefont {H.~T.}\ \bibnamefont
  {Quan}}, \bibinfo {author} {\bibfnamefont {Y.~D.}\ \bibnamefont {Wang}},
  \bibinfo {author} {\bibfnamefont {Y.-X.}\ \bibnamefont {Liu}}, \bibinfo
  {author} {\bibfnamefont {C.~P.}\ \bibnamefont {Sun}}, \ and\ \bibinfo
  {author} {\bibfnamefont {F.}~\bibnamefont {Nori}},\ }\Doi
  {10.1103/PhysRevLett.97.180402} {\bibfield  {journal} {\bibinfo  {journal}
  {Phys. Rev. Lett.},\ }\textbf {\bibinfo {volume} {97}},\ \bibinfo {pages}
  {180402} (\bibinfo {year} {2006})}\BibitemShut {NoStop}%
\bibitem [{\citenamefont {Quan}\ \emph {et~al.}(2007)\citenamefont {Quan},
  \citenamefont {Liu}, \citenamefont {Sun},\ and\ \citenamefont
  {Nori}}]{Quan07}%
  \BibitemOpen
  \bibfield  {author} {\bibinfo {author} {\bibfnamefont {H.~T.}\ \bibnamefont
  {Quan}}, \bibinfo {author} {\bibfnamefont {Y.-x.}\ \bibnamefont {Liu}},
  \bibinfo {author} {\bibfnamefont {C.~P.}\ \bibnamefont {Sun}}, \ and\
  \bibinfo {author} {\bibfnamefont {F.}~\bibnamefont {Nori}},\ }\Doi
  {10.1103/PhysRevE.76.031105} {\bibfield  {journal} {\bibinfo  {journal}
  {Phys. Rev. E},\ }\textbf {\bibinfo {volume} {76}},\ \bibinfo {pages}
  {031105} (\bibinfo {year} {2007})}\BibitemShut {NoStop}%
\bibitem [{\citenamefont {Maruyama}\ \emph {et~al.}(2009)\citenamefont
  {Maruyama}, \citenamefont {Nori},\ and\ \citenamefont {Vedral}}]{Maruyama09}%
  \BibitemOpen
  \bibfield  {author} {\bibinfo {author} {\bibfnamefont {K.}~\bibnamefont
  {Maruyama}}, \bibinfo {author} {\bibfnamefont {F.}~\bibnamefont {Nori}}, \
  and\ \bibinfo {author} {\bibfnamefont {V.}~\bibnamefont {Vedral}},\ }\Doi
  {10.1103/RevModPhys.81.1} {\bibfield  {journal} {\bibinfo  {journal} {Rev.
  Mod. Phys.},\ }\textbf {\bibinfo {volume} {81}},\ \bibinfo {pages} {1}
  (\bibinfo {year} {2009})}\BibitemShut {NoStop}%
\bibitem [{\citenamefont {Rigol}\ \emph {et~al.}(2007)\citenamefont {Rigol},
  \citenamefont {Dunjko}, \citenamefont {Yurovsky},\ and\ \citenamefont
  {Olshanii}}]{Rigol07}%
  \BibitemOpen
  \bibfield  {author} {\bibinfo {author} {\bibfnamefont {M.}~\bibnamefont
  {Rigol}}, \bibinfo {author} {\bibfnamefont {V.}~\bibnamefont {Dunjko}},
  \bibinfo {author} {\bibfnamefont {V.}~\bibnamefont {Yurovsky}}, \ and\
  \bibinfo {author} {\bibfnamefont {M.}~\bibnamefont {Olshanii}},\ }\Doi
  {10.1103/PhysRevLett.98.050405} {\bibfield  {journal} {\bibinfo  {journal}
  {Phys. Rev. Lett.},\ }\textbf {\bibinfo {volume} {98}},\ \bibinfo {pages}
  {050405} (\bibinfo {year} {2007})}\BibitemShut {NoStop}%
\bibitem [{\citenamefont {Yuan}\ \emph {et~al.}(2007)\citenamefont {Yuan},
  \citenamefont {Katsnelson},\ and\ \citenamefont {De~Raedt}}]{Yuan07}%
  \BibitemOpen
  \bibfield  {author} {\bibinfo {author} {\bibfnamefont {S.}~\bibnamefont
  {Yuan}}, \bibinfo {author} {\bibfnamefont {M.~I.}\ \bibnamefont
  {Katsnelson}}, \ and\ \bibinfo {author} {\bibfnamefont {H.}~\bibnamefont
  {De~Raedt}},\ }\Doi {10.1103/PhysRevA.75.052109} {\bibfield  {journal}
  {\bibinfo  {journal} {Phys. Rev. A},\ }\textbf {\bibinfo {volume} {75}},\
  \bibinfo {pages} {052109} (\bibinfo {year} {2007})}\BibitemShut {NoStop}%
\bibitem [{\citenamefont {Cramer}\ \emph
  {et~al.}(2008){\natexlab{a}}\citenamefont {Cramer}, \citenamefont {Dawson},
  \citenamefont {Eisert},\ and\ \citenamefont {Osborne}}]{Cramer08a}%
  \BibitemOpen
  \bibfield  {author} {\bibinfo {author} {\bibfnamefont {M.}~\bibnamefont
  {Cramer}}, \bibinfo {author} {\bibfnamefont {C.~M.}\ \bibnamefont {Dawson}},
  \bibinfo {author} {\bibfnamefont {J.}~\bibnamefont {Eisert}}, \ and\ \bibinfo
  {author} {\bibfnamefont {T.~J.}\ \bibnamefont {Osborne}},\ }\Doi
  {10.1103/PhysRevLett.100.030602} {\bibfield  {journal} {\bibinfo  {journal}
  {Phys. Rev. Lett.},\ }\textbf {\bibinfo {volume} {100}},\ \bibinfo {pages}
  {030602} (\bibinfo {year} {2008}{\natexlab{a}})}\BibitemShut {NoStop}%
\bibitem [{\citenamefont {Cramer}\ \emph
  {et~al.}(2008){\natexlab{b}}\citenamefont {Cramer}, \citenamefont {Flesch},
  \citenamefont {McCulloch}, \citenamefont {Schollw\"ock},\ and\ \citenamefont
  {Eisert}}]{Cramer08b}%
  \BibitemOpen
  \bibfield  {author} {\bibinfo {author} {\bibfnamefont {M.}~\bibnamefont
  {Cramer}}, \bibinfo {author} {\bibfnamefont {A.}~\bibnamefont {Flesch}},
  \bibinfo {author} {\bibfnamefont {I.~P.}\ \bibnamefont {McCulloch}}, \bibinfo
  {author} {\bibfnamefont {U.}~\bibnamefont {Schollw\"ock}}, \ and\ \bibinfo
  {author} {\bibfnamefont {J.}~\bibnamefont {Eisert}},\ }\Doi
  {10.1103/PhysRevLett.101.063001} {\bibfield  {journal} {\bibinfo  {journal}
  {Phys. Rev. Lett.},\ }\textbf {\bibinfo {volume} {101}},\ \bibinfo {pages}
  {063001} (\bibinfo {year} {2008}{\natexlab{b}})}\BibitemShut {NoStop}%
\bibitem [{\citenamefont {Camalet}(2008)}]{Camalet08}%
  \BibitemOpen
  \bibfield  {author} {\bibinfo {author} {\bibfnamefont {S.}~\bibnamefont
  {Camalet}},\ }\Doi {10.1103/PhysRevLett.100.180401} {\bibfield  {journal}
  {\bibinfo  {journal} {Phys. Rev. Lett.},\ }\textbf {\bibinfo {volume}
  {100}},\ \bibinfo {pages} {180401} (\bibinfo {year} {2008})}\BibitemShut
  {NoStop}%
\bibitem [{\citenamefont {Reimann}(2008)}]{Reimann08}%
  \BibitemOpen
  \bibfield  {author} {\bibinfo {author} {\bibfnamefont {P.}~\bibnamefont
  {Reimann}},\ }\Doi {10.1103/PhysRevLett.101.190403} {\bibfield  {journal}
  {\bibinfo  {journal} {Phys. Rev. Lett.},\ }\textbf {\bibinfo {volume}
  {101}},\ \bibinfo {pages} {190403} (\bibinfo {year} {2008})}\BibitemShut
  {NoStop}%
\bibitem [{\citenamefont {Rigol}\ \emph {et~al.}(2008)\citenamefont {Rigol},
  \citenamefont {Dunjko},\ and\ \citenamefont {Olshanii}}]{Rigol08}%
  \BibitemOpen
  \bibfield  {author} {\bibinfo {author} {\bibfnamefont {M.}~\bibnamefont
  {Rigol}}, \bibinfo {author} {\bibfnamefont {V.}~\bibnamefont {Dunjko}}, \
  and\ \bibinfo {author} {\bibfnamefont {M.}~\bibnamefont {Olshanii}},\ }\Doi
  {10.1038/nature06838} {\bibfield  {journal} {\bibinfo  {journal} {Nature
  (London)},\ }\textbf {\bibinfo {volume} {452}},\ \bibinfo {pages} {854}
  (\bibinfo {year} {2008})}\BibitemShut {NoStop}%
\bibitem [{\citenamefont {Garc\'\i{}a-Saez}\ \emph {et~al.}(2009)\citenamefont
  {Garc\'\i{}a-Saez}, \citenamefont {Ferraro},\ and\ \citenamefont
  {Ac\'\i{}n}}]{Gartia09}%
  \BibitemOpen
  \bibfield  {author} {\bibinfo {author} {\bibfnamefont {A.}~\bibnamefont
  {Garc\'\i{}a-Saez}}, \bibinfo {author} {\bibfnamefont {A.}~\bibnamefont
  {Ferraro}}, \ and\ \bibinfo {author} {\bibfnamefont {A.}~\bibnamefont
  {Ac\'\i{}n}},\ }\Doi {10.1103/PhysRevA.79.052340} {\bibfield  {journal}
  {\bibinfo  {journal} {Phys. Rev. A},\ }\textbf {\bibinfo {volume} {79}},\
  \bibinfo {pages} {052340} (\bibinfo {year} {2009})}\BibitemShut {NoStop}%
\bibitem [{\citenamefont {Fine}(2009)}]{Vine09}%
  \BibitemOpen
  \bibfield  {author} {\bibinfo {author} {\bibfnamefont {B.~V.}\ \bibnamefont
  {Fine}},\ }\Doi {10.1103/PhysRevE.80.051130} {\bibfield  {journal} {\bibinfo
  {journal} {Phys. Rev. E},\ }\textbf {\bibinfo {volume} {80}},\ \bibinfo
  {pages} {051130} (\bibinfo {year} {2009})}\BibitemShut {NoStop}%
\bibitem [{\citenamefont {Yuan}\ \emph {et~al.}(2009)\citenamefont {Yuan},
  \citenamefont {Katsnelson},\ and\ \citenamefont {De~Raedt}}]{Yuan09}%
  \BibitemOpen
  \bibfield  {author} {\bibinfo {author} {\bibfnamefont {S.}~\bibnamefont
  {Yuan}}, \bibinfo {author} {\bibfnamefont {M.~I.}\ \bibnamefont
  {Katsnelson}}, \ and\ \bibinfo {author} {\bibfnamefont {H.}~\bibnamefont
  {De~Raedt}},\ }\Doi {10.1143/JPSJ.78.094003} {\bibfield  {journal} {\bibinfo
  {journal} {J. Phys. Soc. Jpn.},\ }\textbf {\bibinfo {volume} {78}},\ \bibinfo
  {eid} {094003} (\bibinfo {year} {2009})}\BibitemShut {NoStop}%
\bibitem [{\citenamefont {Linden}\ \emph {et~al.}(2009)\citenamefont {Linden},
  \citenamefont {Popescu}, \citenamefont {Short},\ and\ \citenamefont
  {Winter}}]{Linden09}%
  \BibitemOpen
  \bibfield  {author} {\bibinfo {author} {\bibfnamefont {N.}~\bibnamefont
  {Linden}}, \bibinfo {author} {\bibfnamefont {S.}~\bibnamefont {Popescu}},
  \bibinfo {author} {\bibfnamefont {A.~J.}\ \bibnamefont {Short}}, \ and\
  \bibinfo {author} {\bibfnamefont {A.}~\bibnamefont {Winter}},\ }\Doi
  {10.1103/PhysRevE.79.061103} {\bibfield  {journal} {\bibinfo  {journal}
  {Phys. Rev. E},\ }\textbf {\bibinfo {volume} {79}},\ \bibinfo {pages}
  {061103} (\bibinfo {year} {2009})}\BibitemShut {NoStop}%
\bibitem [{\citenamefont {Cho}\ and\ \citenamefont {Kim}(2010)}]{Cho10}%
  \BibitemOpen
  \bibfield  {author} {\bibinfo {author} {\bibfnamefont {J.}~\bibnamefont
  {Cho}}\ and\ \bibinfo {author} {\bibfnamefont {M.~S.}\ \bibnamefont {Kim}},\
  }\Doi {10.1103/PhysRevLett.104.170402} {\bibfield  {journal} {\bibinfo
  {journal} {Phys. Rev. Lett.},\ }\textbf {\bibinfo {volume} {104}},\ \bibinfo
  {pages} {170402} (\bibinfo {year} {2010})}\BibitemShut {NoStop}%
\bibitem [{\citenamefont {Schr\"oder}\ and\ \citenamefont
  {Mahler}(2010)}]{Schroder10}%
  \BibitemOpen
  \bibfield  {author} {\bibinfo {author} {\bibfnamefont {H.}~\bibnamefont
  {Schr\"oder}}\ and\ \bibinfo {author} {\bibfnamefont {G.}~\bibnamefont
  {Mahler}},\ }\Doi {10.1103/PhysRevE.81.021118} {\bibfield  {journal}
  {\bibinfo  {journal} {Phys. Rev. E},\ }\textbf {\bibinfo {volume} {81}},\
  \bibinfo {pages} {021118} (\bibinfo {year} {2010})}\BibitemShut {NoStop}%
\bibitem [{\citenamefont {Dong}\ \emph {et~al.}(2007)\citenamefont {Dong},
  \citenamefont {Yang}, \citenamefont {Liu},\ and\ \citenamefont
  {Sun}}]{Dong07}%
  \BibitemOpen
  \bibfield  {author} {\bibinfo {author} {\bibfnamefont {H.}~\bibnamefont
  {Dong}}, \bibinfo {author} {\bibfnamefont {S.}~\bibnamefont {Yang}}, \bibinfo
  {author} {\bibfnamefont {X.~F.}\ \bibnamefont {Liu}}, \ and\ \bibinfo
  {author} {\bibfnamefont {C.~P.}\ \bibnamefont {Sun}},\ }\Doi
  {10.1103/PhysRevA.76.044104} {\bibfield  {journal} {\bibinfo  {journal}
  {Phys. Rev. A},\ }\textbf {\bibinfo {volume} {76}},\ \bibinfo {pages}
  {044104} (\bibinfo {year} {2007})}\BibitemShut {NoStop}%
\bibitem [{\citenamefont {Reimann}(2010)}]{Reimann10}%
  \BibitemOpen
  \bibfield  {author} {\bibinfo {author} {\bibfnamefont {P.}~\bibnamefont
  {Reimann}},\ }\href {http://stacks.iop.org/1367-2630/12/i=5/a=055027}
  {\bibfield  {journal} {\bibinfo  {journal} {New J. Phys.},\ }\textbf
  {\bibinfo {volume} {12}},\ \bibinfo {pages} {055027} (\bibinfo {year}
  {2010})}\BibitemShut {NoStop}%
\end{thebibliography}
\end{document}